\documentclass[10pt,journal,compsoc]{IEEEtran}
\ifCLASSOPTIONcompsoc
  \usepackage[nocompress]{cite}
\else
  \usepackage{cite}
\fi

\ifCLASSINFOpdf
\else
\fi
\usepackage{graphicx}
\usepackage{dcolumn}
\usepackage {subfigure}
\usepackage{bm}
\usepackage{multirow}
\usepackage{enumerate}
\usepackage{textcomp}
\usepackage[letterpaper,hmargin=1in,vmargin=1in]{geometry}

\usepackage{amsmath,amsthm,amssymb,amsfonts,boxedminipage}
\usepackage{algorithm}
\usepackage{algorithmic}
\usepackage[roman]{complexity}

\DeclareMathOperator*{\argmin}{\arg\!\min}
\DeclareMathOperator{\Tr}{Tr}

\newtheorem{theorem}{Theorem}

\newcommand{\abs}[1]{\left\lvert#1\right\rvert}
\newcommand{\norm}[1]{\left\lVert#1\right\rVert}
\newcommand{\ket}[1]{\left|#1 \right\rangle}

\newcommand{\ketbra}[2]{\left| #1 \rangle \langle #2 \right|}

\begin{document}
%
\title{An improved quantum algorithm for ridge regression}

\author{Chao-Hua Yu, Fei Gao and Qiao-Yan Wen 
\IEEEcompsocitemizethanks{\IEEEcompsocthanksitem Chao-Hua Yu is with the State Key Laboratory of Networking and Switching Technology, Beijing University of Posts and Telecommunications, Beijing, 100876, China, and with the School of Physics, University of Western Australia, Crawley, Western Australia 6009, Australia. E-mail: quantum.ych@gmail.com \protect
\IEEEcompsocthanksitem  F. Gao and Q.-Y. Wen are with the State Key Laboratory of Networking and Switching
Technology, Beijing University of Posts and Telecommunications, Beijing, 100876, China. E-mail: gaof@bupt.edu.cn; wqy@bupt.edu.cn.}\protect\\
\thanks{Manuscript received ; revised.}}

\markboth{IEEE TRANSACTIONS ON KNOWLEDGE AND DATA ENGINEERING,~Vol~XX., No.,~XX}%
{Shell \MakeLowercase{\textit{et al.}}: Bare Demo of IEEEtran.cls for Computer Society Journals}
%



\IEEEtitleabstractindextext{%
\begin{abstract}
Ridge regression (RR) is an important machine learning technique which introduces a regularization hyperparameter $\alpha$ to ordinary multiple linear regression for analyzing data suffering from multicollinearity. In this paper, we present a quantum algorithm for RR, where the technique of parallel Hamiltonian simulation to simulate a number of Hermitian matrices in parallel is proposed and used to develop a quantum version of $K$-fold cross-validation approach, which can efficiently estimate the predictive performance of RR. Our algorithm consists of two phases: (1) using quantum $K$-fold cross-validation to efficiently determine a good $\alpha$ with which RR can achieve good predictive performance, and then (2) generating a quantum state encoding the optimal fitting parameters of RR with such $\alpha$, which can be further utilized to predict new data. Since indefinite dense Hamiltonian simulation has been adopted as a key subroutine, our algorithm can efficiently handle non-sparse data matrices. It is shown that our algorithm can achieve exponential speedup over the classical counterpart for (low-rank) data matrices with low condition numbers. But when the condition numbers of data matrices is large to be amenable to full or approximately full ranks of data matrices, only polynomial speedup can be achieved.

\end{abstract}

}

\maketitle

\IEEEdisplaynontitleabstractindextext

\IEEEpeerreviewmaketitle

\IEEEraisesectionheading{\section{Introduction}\label{sec:introduction}}

Dating from the 80's of last century, quantum computing has been shown to be more computationally powerful in solving certain problems than classical computing \cite{QCQI10,QA16,QARM16,AQT18}. In the past decade, it has been brought into the field of machine learning, which is a subfield of computer science and studies how to learn from data and make predictions on new data \cite{ML12}, giving birth to a new disciplinary research field---quantum machine learning. Since its inception, quantum machine learning has become a booming research field attracting worldwide attentions, and a number of efficient quantum algorithms have been proposed for various machine learning tasks \cite{QDM14,SSP15,QML17,DYLL17}.

Linear regression (LR) is one of the most important machine learning tasks with wide applications in many scientific fields including biology, behavioristic, sociology, finance, and so on \cite{ML12}. Given $N$ data points $(\mathbf{x}_i,y_i)_{i=1}^N$, where $\mathbf{x}_i=(x_{i1},\cdots,x_{iM})^T\in \mathbb{R}^M$ is a vector of $M$ independent (exploratory, input) variables and $y_i\in \mathbb{R}$ is the scalar dependent (response, output) variable, LR assumes that $\mathbf{x}_i$ and $y_i$ are linearly correlated and attempts to construct a linear function $f(\mathbf{x})=\mathbf{w}^T\mathbf{x}$ characterized by fitting parameters $\mathbf{w}=(w_1,\cdots,w_M)^T$ that can best fit such relationship, i.e., making every $f(\mathbf{x}_i)$ as close as possible to $y_i$ . It should be emphasized that $\mathbf{x}$ can be generated by a nonlinear map on some original data, such as polynomial function, which enables LR to fit nonlinear function.

The simplest LR model is \textit{ordinary linear regression} (OLR), where the optimal fitting parameters $\mathbf{w}=(\mathbf{X}^T\mathbf{X})^{-1}\mathbf{X}^T\mathbf{y}$ are determined via least squares method of minimizing the sum of squared residuals. Here $\mathbf{y}=(y_1,\cdots,y_N)^T$, and $\mathbf{X}=(\mathbf{x}_1, \cdots, \mathbf{x}_N)^T$ is called \emph{design matrix}. However, OLR in practice is often far from satisfaction \cite{ML12,Ridge70,vW15} when suffering multicollinearity of independent variables of data points (which makes $\mathbf{X}^T\mathbf{X}$ not invertible) or overfitting. These two difficulties substantially restrict the effectiveness of OLR when putting it into real-world applications. To circumvent them, Hoerl et al. \cite{vW15} put forward  a generalized version of OLR---\textit{ridge regression} (RR), in which some regularization of $\mathbf{w}$ is introduced into optimization. This leads to the optimal fitting parameters of RR being $\mathbf{w}=(\mathbf{X}^T\mathbf{X}+\alpha \mathbf{I})^{-1}\mathbf{X}^T\mathbf{y}$, where $\alpha$ denotes regularization hyperparameter and $\mathbf{I}$ is the identity matrix. However, choosing an appropriate $\alpha$ with which RR can achieve the best (or approximately best) predictive performance is of great challenge.

As of now, a series of quantum algorithms for LR have been proposed. By building on the well-known quantum algorithm for solving linear systems of equations proposed by Harrow, Hassidim, and Lloyd (HHL) \cite{HHL09}, Wiebe et al. \cite{WBL12} first provided a quantum algorithm that can efficiently determine the fitting quality of OLR over an exponentially large data set with a sparse design matrix. Their results were later improved and directly extended to tackle RR \cite{LZ17}. Lately, different from the previous algorithms \cite{HHL09,LZ17} which are efficient only for the data sets with sparse design matrices, Schuld et al. provided a quantum algorithm for prediction by OLR that can efficiently process low-rank non-sparse design matrices \cite{SSP16}. More recently, Wang suggested a quantum linear regression algorithm that works in the standard oracle model and can efficiently output the optimal fitting parameters in the classical form \cite{Wang17}. However, with the exception of prior work for quantum RR \cite{LZ17}, almost all of these quantum linear regression algorithms are based on OLR rather than RR, thus cannot combat multicollinearity and overfitting mentioned above.

In this paper, to deeply explore how and to what extent RR can be done by quantum computing faster than by classical computing, we design a more comprehensive quantum algorithm for RR. Inspired by the technique of $K$-fold cross-validation \cite{Ridge70} which has been widely used to evaluate the predictive performance of many machine learning algorithms \cite{ML12,HZZL16}, we propose its quantum version to efficiently evaluate the predictive performance of RR. Our quantum algorithm will use the quantum $K$-fold cross-validation to determine a good $\alpha$ for RR, and then generate a quantum state encoding the fitting parameters of RR with such $\alpha$ in the amplitudes. It is shown that our algorithm is exponentially faster than the classical counterpart, when processing (low-rank) design matrices with relatively small elements and low condition numbers, but when design matrices have large condition numbers amenable to full or approximately full ranks of data matrices, only polynomial speedup can be achieved. Our algorithm improves the existing quantum algorithm for RR, i.e., LZ's algorithm \cite{LZ17}, from two aspects. First, since our algorithm uses indefinite dense Hamiltonian simulation \cite{RSML18} as the key subroutine, our algorithm has no dependence on the sparsity of design matrices, but has slightly worse dependence on the error, whereas LZ's uses sparse Hamiltonian simulation as the subroutine and can only efficiently tackle sparse design matrices. Second, our algorithm presents an efficient procedure, i.e., quantum $K$-fold cross-validation, to determine a good $\alpha$ for RR, while LZ's algorithm does not address this important task.


Just as other HHL-based quantum machine learning algorithms with several caveats \cite{SA15}, our algorithm also faces similar caveats. First, our algorithm assumes that efficient quantum access to the entries of $\mathbf{X}$ and $\mathbf{y}$ is provided. This can be achieved by quantum random access memory (QRAM) \cite{QRAM08}, for which there is no general implementation in quantum hardware to date. However, if the entries can be efficiently computed by simple and explicit formula, the quantum access can be efficiently implemented directly without QRAM. Second, our algorithm does not output the classical form of the optimal fitting parameters $\mathbf{w}$, but a quantum state $\ket{\mathbf{w}}$ encoding $\mathbf{w}$ in its amplitudes. Nevertheless, the state can be further used to efficiently predict new data via swap test \cite{BCWW01,LMR13}. Finally, our algorithm is exponentially fast when the condition number of the design matrix $\mathbf{X}$ is relatively low. The condition number may be reduced by preconditioning $\mathbf{X}$.

\section{Review of ridge regression}
\label{sec:ClassRRReview}
Given a set of $N$ data points $(\mathbf{x}_i,y_i)_{i=1}^{N}$ as described above, RR aims at finding a linear function $f(\mathbf{x})=\mathbf{x}^T\mathbf{w}=\sum_{j=1}^M x_jw_j$ characterized by the fitting parameters $\mathbf{w}=(w_1,\cdots,w_M)^T$ that makes all $f(\mathbf{x}_i)$ as close as possible to $y_i$ \cite{ML12,vW15,Ridge70}. Different from the OLR where the sum of squared residuals is minimized,  RR minimizes the sum of squared residuals plus a fraction of regularization of $\mathbf{w}$ and has the optimal fitting parameters
\begin{eqnarray}
\mathbf{w} &=&\argmin_{\mathbf{w}}\sum_{i=1}^N |f(\mathbf{x}_i)-y_i|^2+\alpha\norm{\mathbf{w}}^2 \nonumber\\
&=&(\mathbf{X}^T\mathbf{X}+\alpha \mathbf{I})^{-1}\mathbf{X}^T\mathbf{y},
\label{OptimalW}
\end{eqnarray}
where $\norm{\mathbf{v}}$ is the 2-norm of any vector $\mathbf{v}$. Evidently, OLR is a special case of RR with $\alpha=0$. Write $\mathbf{X}$ in the reduced singular value decomposition \cite{HLA06} form $\mathbf{X}=\sum_{j=1}^R\lambda_j|\mathbf{u}_j\rangle\langle\mathbf{v}_j|$, where $R$ is the rank of $\mathbf{X}$,  $\lambda_j$ are the nonzero singular values, and $|\mathbf{u}_j\rangle$ ($|\mathbf{v}_j\rangle$) are the corresponding left (right) normalized singular vectors. Adding another $N-R$ normalized vectors $\ket{\mathbf{u}_{R+1}},\cdots,\ket{\mathbf{u}_N}$ that make $\ket{\mathbf{u}_{1}},\cdots,\ket{\mathbf{u}_N}$ become an orthonormal basis spanning the whole space $\mathbb{R}^N$, $\mathbf{y}/\norm{\mathbf{y}}$ can be written as a linear combination of $\{|\mathbf{u}_j\rangle\}_1^N$, $\mathbf{y}/\norm{\mathbf{y}}=\sum_{j=1}^N \beta_j |\mathbf{u}_j\rangle$ with $\sum_{j=1}^N \beta_j^2=1$, and thus $\mathbf{w}$ can be rephrased as
\begin{eqnarray}
\label{SVDOptimalW}
\mathbf{w}=\sum_{j=1}^R \frac{\lambda_j}{\lambda_j^2+\alpha}\beta_j\norm{\mathbf{y}}|\mathbf{v}_j\rangle,
\end{eqnarray}
which depends on the choice of $\alpha$. After attaining $\mathbf{w}$, one can predict the output $\tilde{y}$ of any new input $\tilde{\mathbf{x}}$ via computing $\tilde{y}=\mathbf{w}^T\tilde{\mathbf{x}}$.  So the predictive squared error sum for all the training data points is
\begin{eqnarray*}
&&\norm{X\mathbf{w}-\mathbf{y}}^2 \nonumber\\
&=&\norm{\mathbf{y}}^2\left(\sum_{j=1}^R\left(1-\frac{\lambda_j^2}{\lambda_j^2+\alpha}\right)^2\beta_j^2+\sum_{j=R+1}^N\beta_j^2\right) \nonumber\\
&\geq& \norm{\mathbf{y}}^2\left(1-\Lambda(2-\Lambda)\left(\sum_{j=1}^R\beta_j^2\right)\right),
\end{eqnarray*}
since $\sum_{j=R+1}^N\beta_j^2=1-(\sum_{j=1}^R\beta_j^2)$, where $\Lambda=\max_{j=1,\cdots,R}\frac{\lambda_j^2}{\lambda_j^2+\alpha}$ and $0<\Lambda <1$. If $\sum_{j=1}^R\beta_j^2$ is small, the error sum would be very large, meaning that the model is badly constructed; otherwise, the error sum is small. As a result, when the RR model is well constructed, the support of $\mathbf{y}/\norm{\mathbf{y}}$ in the space spanned by $\{\ket{\mathbf{u}_j}\}_{j=1}^R$, i.e., $\sum_{j=1}^R \beta_j^2$, should be large to be close to 1.

Therefore, it is of great importance to choose a good $\alpha$ so that RR with such $\alpha$ can achieve good predictive performance, and then to obtain the $\mathbf{w}$ of RR with such $\alpha$.

\section{Quantum algorithm}
In the following, we design a quantum algorithm for RR. It consists of two subroutines: a quantum algorithm for generating the quantum state encoding the optimal fitting parameters $\mathbf{w}$ (Eqs.~\eqref{OptimalW} and \eqref{SVDOptimalW}), and a quantum algorithm for finding a good $\alpha$. Throughout the algorithm, we assume we are provided the quantum oracles
$$O_\mathbf{X}:|j\rangle|k\rangle|0\rangle \mapsto |j\rangle|k\rangle|x_{jk}\rangle$$
and
$$O_\mathbf{y}: |j\rangle|0\rangle \mapsto |j\rangle|y_j\rangle,$$
which can efficiently access the entries of $\mathbf{X}$ and $\mathbf{y}$ in time $O(\polylog(MN))$ and $O(\polylog(N))$, respectively. This holds when the entries of $\mathbf{X}$ and $\mathbf{y}$ are efficiently computable or are stored in QRAM \cite{QRAM08}. In general, $\mathbf{X}$ is not too much skewed, and $\norm{\mathbf{X}}_{\max}$ and $\norm{\mathbf{y}}_{\max}$ are not too large, hence we assume $M=\Theta(N)$ and $\norm{\mathbf{X}}_{\max},\norm{\mathbf{y}}_{\max}=\Theta(1)$ hereafter.

\subsection{Algorithm 1: generating a quantum state encoding the optimal fitting parameters}
We first give a quantum algorithm to generate a quantum state $\ket{\mathbf{w}}$ that approximates the normalized $\mathbf{w}$ within error $\epsilon$. From Eq.~\eqref{SVDOptimalW}, it is easy to see that, to obtain $\mathbf{w}$, we need perform singular value decomposition on $\mathbf{X}$. To achieve this, the recently invented technique of indefinite non-sparse Hamiltonian simulation \cite{RSML18} is adopted. Given a Hermitian matrix $A\in \mathbb{C}^{N\times N}$ and efficient quantum access to its entries, by embedding $A$ into a larger one-sparse Hermitian matrix, it is able to simulate the unitary matrix $e^{\frac{-iAt}{N}}$ for time $t$ within error $\epsilon$ in time $O\left(\polylog (N) t^2\|A\|_{\max}^2/\epsilon\right)$, where $\norm{A}_{\max}:=\max\limits_{ij} |A_{ij}|$. However, in our problem, since $\mathbf{X}$ is generally not Hermitian, we extend it to a larger but Hermitian matrix
\begin{eqnarray}
\tilde{\mathbf{X}}=\left[
\begin{matrix}
0 &\mathbf{X} \\
\mathbf{X}^T &0
\end{matrix}
\right]\in \mathbb{R}^{(N+M)\times(N+M)},
\end{eqnarray}
which is of $2R$ nonzero eigenvalues $\{\pm\lambda_j\}_{j=1}^R$ and corresponding normalized eigenvectors $\{\ket{\mathbf{u}_j,\pm \mathbf{v}_j}:= \left(|0,\mathbf{u}_j\rangle\pm|1,\mathbf{v}_j\rangle\right)/\sqrt{2}\in \mathbb{R}^{N+M}\}_{j=1}^R$, where
\begin{eqnarray}
|0,\mathbf{u}_j\rangle=\left[
\begin{matrix}
|\mathbf{u}_j\rangle\\
\mathbf{0}
\end{matrix}
\right],
|1,\mathbf{v}_j\rangle=\left[
\begin{matrix}
\mathbf{0}\\
|\mathbf{v}_j\rangle
\end{matrix}
\right].
\end{eqnarray}
Without loss of generality, we assume $\frac{\lambda_j}{N+M}\in \left[1/\kappa,1\right]$, where $\kappa$ is the condition number of $\mathbf{X}$. In addition, from Eq.~\eqref{SVDOptimalW}, it is easy to see that too small $\alpha$ will make RR reduced to OLR and too large $\alpha$ will make the optimal fitting parameters approach zero, thus we choose $\alpha$ satisfying $\Theta\left(\frac{(N+M)^2}{\kappa^2}\right)\leq \alpha \leq \Theta\left((N+M)^2\right)$.

\begin{figure}[hbt]
\begin{center}
\includegraphics[scale=0.36]{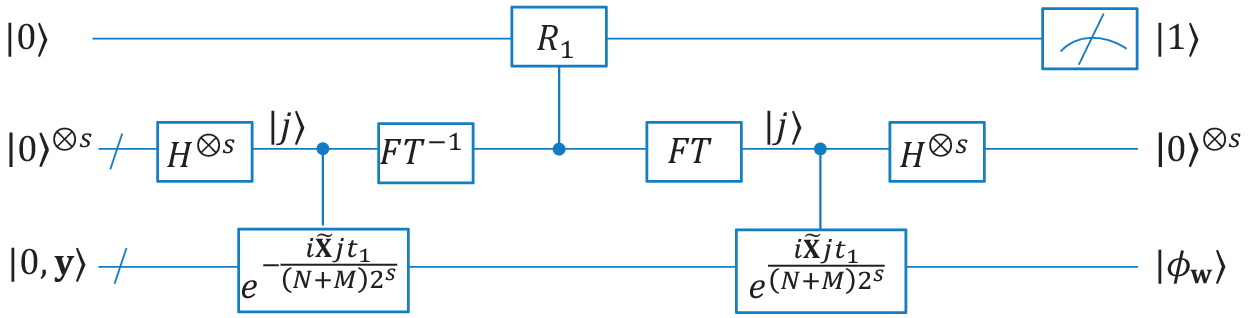}
\end{center}
\caption{\label{fig1} Quantum circuit for algorithm 1. Here the '$/$' denotes a bundle of
wires, $H$ denotes the Hadamard operation, $FT$ represents the quantum Fourier transformation and $FT^{-1}$ is its inverse \cite{QCQI10}, and $s$ is the number of qubits for estimating the eigenvalues of $\frac{\tilde{\mathbf{X}}}{N+M}$ in step (2), controlled-$R_1$ denotes controlled rotation in step (3).}
\end{figure}

The first algorithm proceeds as following steps and the schematic is given in Fig. \ref{fig1}:

(1) Prepare the $(N+M)$-dimensional quantum state $|0,\mathbf{y}\rangle =\left(|\mathbf{y}\rangle^T, \mathbf{0}\right)^T=\sum_{j=1}^N \beta_j\ket{0,\mathbf{u}_j}$ by directly expanding the state $|\mathbf{y}\rangle:=\mathbf{y}/\norm{\mathbf{y}}$.

Here we assume $|\mathbf{y}\rangle$ can be generated efficiently in time $O(\polylog(N))$. As shown in appendix \ref{appendix:StatePre},  when $\mathbf{y}$ is \textit{balanced} \cite{Wang17} in the sense that
$$\frac{\sum_{j=1}^N|y_{j}|^2}{N\|\mathbf{y}\|_{\max}^2}=\Omega(1),$$
$|\mathbf{y}\rangle$ can be efficiently generated in time $O\left(\polylog N\right)$ via $O_\mathbf{y}$. Alternatively, $|\mathbf{y}\rangle$ can also be efficiently prepared when for any $i_1,i_2$, $\sum_{i=i_1}^{i_2}\abs{y_i}^2$ are efficiently computable \cite{GR02}.


(2) Add another register in the state $|0\cdots 0\rangle$ to the above state $\ket{0,\mathbf{y}}$, and perform phase estimation by simulating $e^{\frac{-i\tilde{\mathbf{X}}t_1}{N+M}}$ \cite{RSML18} for some evolution time $t_1$ to reveal the eigenvalues and eigenvectors of $\frac{\tilde{\mathbf{X}}}{N+M}$ as
\begin{eqnarray}
\label{alg1:PE}
\sum_{j=1}^R \beta_j |\mathbf{u}_j,\pm \mathbf{v}_j\rangle\ket{\frac{\pm \lambda_j}{N+M}}/\sqrt{2}.
\end{eqnarray}

Here for convenience we assume $|\mathbf{y}\rangle$ fully lies in the subspace $\{|\mathbf{u}_j\rangle\}_{j=1}^R$, namely $|\mathbf{y}\rangle=\sum_{j=1}^R \beta_j |\mathbf{u}_j\rangle$ with $\sum_{j=1}^R \beta_j^2=1$ and thus $|0,\mathbf{y}\rangle=\sum_{j=1}^R \beta_j |0,\mathbf{u}_j\rangle=\sum_{j=1}^R \beta_j |\mathbf{u}_j,\pm \mathbf{v}_j\rangle/\sqrt{2}$. If $|\mathbf{y}\rangle$, more generally, does not fully lie in the subspace $\{|\mathbf{u}_j\rangle\}_{j=1}^R$, the state of Eq.~\eqref{alg1:PE} would be
\begin{align}
\label{alg1:PEComplete}
\sum_{j=1}^R \beta_j |\mathbf{u}_j,\pm \mathbf{v}_j\rangle\ket{\frac{\pm \lambda_j}{N+M}}/\sqrt{2}\nonumber\\
+\sum_{j=R+1}^N \beta_j \ket{0,\mathbf{u}_j}\ket{0\cdots 0}.
\end{align}
Nonetheless, we can efficiently transform the state of Eq.~\eqref{alg1:PEComplete} to the state of Eq.~\eqref{alg1:PE}: introduce another qubit $\ket{0}$ and rotate it to $\ket{1}$ when the eigenvalue stored in the second register (eigenvalue register) is nonzero, and then measure the qubit to see the outcome $\ket{1}$. The measurement probability is $\sum_{j=1}^R \beta_j^2 \approx 1$ as discussed in Sec.\ref{sec:ClassRRReview}, so the transformation is quite efficient. After successful measurement, we obtain the state of Eq.~\eqref{alg1:PE} by letting $\beta_j\leftarrow \beta_j/\sqrt{\sum_{j=1}^R \beta_j^2}$ for $j=1,\cdots,R$.


(3) Add one qubit and rotate it from $|0\rangle$ to $\sqrt{1-C_1^2h^2(\pm\lambda_j,\alpha)}|0\rangle + C_1h(\pm\lambda_j, \alpha)|1\rangle$ controlled on $|\frac{\pm \lambda_j}{N+M}\rangle$, where $h(\lambda,\alpha):=\frac{(N+M)\lambda}{\lambda^2+\alpha}$ and $C_1=O\left(\max_{\lambda_j}h(\lambda_j,\alpha)\right)^{-1}=O(1/\kappa)$. As shown in appendix \ref{appendix:MaxOfh}, the maximum of $h(\lambda_j,\alpha)$ as well as $C_1$ depends on the actual choice of $\alpha$, but $C_1h(\lambda_j,\alpha)=\Omega(1/\kappa)$ for all possible $\alpha$. Then we undo phase estimation and obtain
\begin{eqnarray}
\sum_{j=1}^R\beta_j|\mathbf{u}_j,\pm \mathbf{v}_j\rangle \Big( \sqrt{1-C_1^2h^2(\pm \lambda_j,\alpha)}|0\rangle  \nonumber\\
+C_1h(\pm\lambda_j,\alpha)|1\rangle \Big).
\end{eqnarray}

(4) Measure the last qubit to get $|1\rangle$ and project the first register onto the $\mathbf{v}_j$ part. The final state of the first register approximates
\begin{eqnarray}
|\phi_\mathbf{w}\rangle:=\frac{\sum_{j=1}^RC_1\beta_jh\left(\lambda_j,\alpha\right)|\mathbf{v}_j\rangle}{\sqrt{\sum_{j=1}^RC_1^2\beta_j^{2}h^2\left(\lambda_j,\alpha\right)}}
\propto \mathbf{w},
\label{alg1:DesState}
\end{eqnarray}
which is proportional to Eqs.~\eqref{OptimalW} and \eqref{SVDOptimalW} as desired. The success probability of getting $|1\rangle$ is $\sum_{j=1}^R C_1^2\beta_j^{2}h^2\left(\lambda_j,\alpha\right)=\Omega(1/\kappa^2)$, which implies that $O(\kappa^2)$ repetitions are enough to yield the desirable state with a large probability, and this can be improved by amplitude amplification \cite{AA02} with $O(\kappa)$ repetitions.

Given the quantum form $|\tilde{\mathbf{x}}\rangle$ of a new input data $\tilde{\mathbf{x}}$, the state $|\phi_\mathbf{w}\rangle$ can be used to predict the output $\tilde{\mathbf{y}}=\mathbf{w}^T\tilde{\mathbf{x}}$ up to some factor by evaluating the inner product of $|\tilde{\mathbf{x}}\rangle$ and  $|\phi_\mathbf{w}\rangle$  via swap test \cite{BCWW01,LMR13}.

\subsection{Time complexity of algorithm 1.}
The time complexity of algorithm 1 is dominated by phase estimation and amplitude amplification. In step (2), the eigenvalues $\pm \frac{\lambda_j}{N+M}\in \pm\left[1/\kappa,1\right]$ are estimated within error $O(1/t)$ via phase estimation. Consequently, the relative error of estimating $h(\pm\lambda_j,\alpha)$ scales as $O(\kappa/t)$ no matter how $\alpha$ is chosen, but its actual scale depends on $\alpha$ as detailed in appendix \ref{appendix:MaxOfh}. Therefore, $t_1=O(\kappa/\epsilon)$ is taken to ensure the final state approximates $|\phi_\mathbf{w}\rangle$  within error $\epsilon$. Thus, according to \cite{RSML18}, phase estimation takes time $O(\norm{\mathbf{X}}_{\max}^2\polylog(N+M)\kappa^2/\epsilon^3)$. Considering amplitude amplification of $O(\kappa)$ repetitions in step (4), it takes total time $O(\norm{\mathbf{X}}_{\max}^2\polylog(N+M)\kappa^3/\epsilon^3)$ to generate $|\phi_\mathbf{w}\rangle$. Since $\Tr(\mathbf{X}^T\mathbf{X})=\sum_{j=1}^R \lambda_j^2=\sum_{ij}x_{ij}^2 \leq NM \norm{\mathbf{X}}_{\max}^2$ and $\lambda_j \in [\frac{N+M}{\kappa},N+M]$, we have $\frac{R}{\kappa^2} \leq \frac{NM\norm{X}_{\max}^2}{(N+M)^2}$, and thus the rank $R$ of $\mathbf{X}$ is upper bounded as $R=O(\kappa^2)$ due to $M=\Theta(N)$ and $\norm{X}_{\max}=\Theta(1)$.

The best known classical RR algorithm has time complexity $O\left(NM+N^2R\log(\frac{R}{\epsilon})/\epsilon^2\right)$ \cite{CLLKZ15}. Under the assumptions that $M=\Theta(N)$ and $\norm{\mathbf{X}}_{\max}=\Theta(1)$, and letting $1/\epsilon=O(\polylog N)$, our algorithm 1 takes time $O(\polylog(N)\kappa^3)$, while the classical algorithm takes time $\tilde{O}(\polylog(N)N^2R)$, where $\tilde{O}$ is used to suppress the relatively small quantity of $\log(R/\epsilon)$. When $\kappa$ is large with $\kappa=O(\sqrt{N})$ which is amenable to full or approximately full rank of $\mathbf{X}$ (i.e., $R=O(N)$), algorithm 1 achieves (approximately) quadratic speedup over the classical algorithm, because in this case algorithm 1 has time complexity $O(\polylog(N)N^{3/2})$, whereas the classical algorithm has time complexity $\tilde{O}(\polylog(N)N^3)$. However, when $\kappa$ is small with $\kappa=\polylog(N)$ which implies low rank of $\mathbf{X}$ ($R=\polylog(N)$), algorithm 1 has time complexity $O(\polylog(N))$, while the classical algorithm has time complexity $\tilde{O}(\polylog(N)N^2)$, so our algorithm 1 in this case is exponentially faster than the classical algorithm.

Compared with LZ's quantum RR algorithm \cite{LZ17} whose time complexity is $O\left(\log(N+M)s^2\kappa_R^3/\epsilon^2\right)$, where $s$ is the sparsity of design matrix and $\kappa_R=\frac{\max\{1,\frac{\sqrt{\alpha}}{N+M}\}}{\min\{1/\kappa,\frac{\sqrt{\alpha}}{N+M}\}}=O(\kappa)$ (Note that the singular values of design matrix in our algorithm are assumed to be in $[(N+M)/\kappa,N+M]$, while those in LZ's algorithm are assumed to be in $[1/\kappa,1]$), the time complexity of our algorithm 1 has the same dependence on $\kappa$ as LZ's result, whereas the dependence on $\epsilon$ is worse by a factor $\epsilon^{-1}$. However, our algorithm 1 has no dependence on $s$, which makes it capable to efficiently handle non-sparse design matrices, and is exponentially faster than LZ's algorithm for non-sparse design matrices with $s=O(N)$ when $\kappa, 1/\epsilon=O(\polylog N)$.

Just as LZ's algorithm, our algorithm 1 can also output the scale parameter $\norm{\mathbf{w}}^2$ which tells how $\mathbf{w}$ is rescaled to $\ket{\mathbf{w}}$. From Eq.~\eqref{SVDOptimalW}, we can see that
\begin{eqnarray}
\norm{\mathbf{w}}^2&=&\sum_{j=1}^R \left(\frac{\lambda_j}{\lambda_j^2+\alpha}\right)^2\beta_j^2\norm{\mathbf{y}}^2 \nonumber\\
&=&\sum_{j=1}^R \frac{h^2(\lambda_j,\alpha)\beta_j^2\norm{\mathbf{y}}^2}{(N+M)^2} \nonumber\\
&=&\frac{P\norm{\mathbf{y}}^2}{C_1^2(N+M)^2},
\end{eqnarray}
where $P=\sum_{j=1}^R C_1^2\beta_j^{2}h^2\left(\lambda_j,\alpha\right)=\Omega(1/\kappa^2)$ represents the measurement probability in step (4) of algorithm 1.
Just as estimating $\norm{\mathbf{y}}^2$ shown in appendix \ref{appendix:StatePre}, $P$ (as well as $\norm{\mathbf{w}}^2$) is estimated via amplitude estimation \cite{AA02} within relative error $\epsilon$ with $O\left(\sqrt{\frac{1-P}{P}}\frac{1}{\epsilon}\right)=O\left(\frac{\kappa}{\epsilon}\right)$ repetitions of steps (1)-(3). Since the runtime of each repetition is dominated by that of phase estimation as discussed above, i.e., $O(\norm{\mathbf{X}}_{\max}^2\polylog(N+M)\kappa^2/\epsilon^3)$, $\norm{\mathbf{w}}^2$ can be estimated within relative error $\epsilon$ in time $O(\norm{\mathbf{X}}_{\max}^2\polylog(N+M)\kappa^3/\epsilon^4)$. Considering $M=\Theta(N)$ and $\norm{X}_{\max}=\Theta(1)$, this procedure would be efficient when $\kappa,1/\epsilon=\polylog(N)$.

\subsection{Algorithm 2: choosing a good $\alpha$.}

Choosing a good value of $\alpha$ which allows the prediction of future data is a critical part of RR. A common and efficient method for choosing a good $\alpha$ is to choose the best one out of a number of candidate $\alpha$'s, so that RR with such $\alpha$ has the best predictive performance \cite{vW15}. The most common method for evaluating the predictive performance of RR as well as other linear regression tasks is $K$-fold cross-validation \cite{vW15}. Let us outline how to combine these two methods to determine the best $\alpha$. First, the set of $N$ data points is divided into $K$ ($2 \leq K\leq N$) subsets and the $l$-th ($l=1,\cdots,K$) subset contains the data points $(\mathbf{x}_j,y_j)$ with $j \in S_l$, where
\begin{eqnarray}
S_l:=\{(l-1)N/K+1,\cdots,lN/K\}
\label{eq:10}
\end{eqnarray}
is used to mark the numbers of data points assigned to the $l$-th subset. Then $K$ turns of training-test procedures are run, where in the $l$-th turn the $l$-th subset is taken as the test set and the others are taken as the training set. After that, a measure is used to evaluate the predictive performance of this model for a certain $\alpha$. The $\alpha$ over all the candidates corresponding to the best predictive performance is chosen as the final $\alpha$. The details are shown as follows.

Let $\mathbf{X}_l\in \mathbb{R}^{N/K\times M}$ be the matrix containing the rows $S_l$ of $\mathbf{X}$ which corresponds to the $l$-th subset, and $\mathbf{X}_{-l}\in \mathbb{R}^{N\times M}$ be the matrix $\mathbf{X}$ but replacing the elements in the rows of $S_l$ with zeros. Evidently, the rank of $\mathbf{X}_{-l}$ is equal or less than that of $\mathbf{X}$. $\mathbf{X}_{-l}$ can be written in the singular value decomposition $\mathbf{X}_{-l}=\sum_{j=1}^{R_l}\lambda_{lj}|\mathbf{u}_{lj}\rangle\langle \mathbf{v}_{lj}|$, where $\lambda_{lj}$ are its singular values, $|\mathbf{u}_{lj}\rangle$ ($|\mathbf{v}_{lj}\rangle$) are their corresponding left (right) singular vectors, $R_l$ is its rank and $\leq R$ obviously. All $\lambda_{lj}$ lie in $\left(\frac{N+M}{\kappa'}, N+M\right)$, and $\kappa'= O(\kappa)$ by taking $K=\Omega\left(\frac{NM\norm{\mathbf{X}}_{\max}^2\kappa^2}{(N+M)^2}\right)$, a good example of choosing such $K$ being leave-one-out cross-validation; see appendix \ref{appendix:ScaSingXl} for more details on the scale of $\lambda_{lj}$. Similarly, we define $\mathbf{y}_l$ and $\mathbf{y}_{-l}$.

In the $l$-th turn, according to the Eq.~\eqref{SVDOptimalW}, the optimal fitting parameters are
\begin{eqnarray}
\mathbf{w}_l&=&(\mathbf{X}_{-l}^T\mathbf{X}_{-l}+\alpha \mathbf{I})^{-1}\mathbf{X}_{-l}^T\mathbf{y}_{-l}.
\label{alg2:DesClaL}
\end{eqnarray}
Consequently, the squared residual sum of prediction of $l$-th turn is $\|\mathbf{y}_l-\mathbf{X}_l\mathbf{w}_l\|^2$, and the total squared residual sum over $K$ turns is
\begin{align}
&\sum_{l=1}^K\|\mathbf{y}_l-\mathbf{X}_l\mathbf{w}_l\|^2 \nonumber\\
=&\sum_{l=1}^K (\norm{\mathbf{y}_l}^2+\norm{\mathbf{X}_l\mathbf{w}_l}^2-2\mathbf{y}_l^T\mathbf{X}_l\mathbf{w}_l)\nonumber\\
=&E_1(\alpha)+E_2(\alpha)+E_3(\alpha).
\label{eq:SRS}
\end{align}
Based on the sum, we define the predictive error function $E(\alpha)$ as
\begin{align}
\label{alg2:Ealpha}
E(\alpha)&=\frac{\sum_{l=1}^K\|\mathbf{y}_l-\mathbf{X}_l\mathbf{w}_l\|^2}{\sum_{l=1}^K (\norm{\mathbf{y}_l}^2+\norm{\mathbf{X}_l\mathbf{w}_l}^2)} \nonumber\\
&=1+\frac{E_3(\alpha)}{E_1(\alpha)+E_2(\alpha)},
\end{align}
which is a scale-invariant measure of the predictive performance of RR with a certain $\alpha$. The smaller $E(\alpha)$ is, the better the predictive performance is. If $\mathbf{y}_l$ and $\mathbf{X}_l\mathbf{w}_l$ are close, then $E(\alpha)$ is close to zero; otherwise, if they are far apart, it is close to one. Therefore, we can choose the best $\alpha$ by minimizing $E(\alpha)$ over all candidate $\alpha$'s. Given a set of candidate $\alpha$'s, $\{\alpha_1,\cdots,\alpha_L\}$, our objective is to choose $\widehat{\alpha}$ such that
\begin{eqnarray}
\widehat{\alpha}=\argmin_{\alpha \in \{\alpha_1,\cdots,\alpha_L\}}E(\alpha).
\end{eqnarray}
Normally, we uniformly take these $L$ candidate values of $\alpha$ in the prespecified range $[\alpha_{\min}=\Theta\left(\frac{(N+M)^2}{\kappa^2}\right),\alpha_{\max}=\Theta\left((N+M)^2\right)]$, e.g., $\left[\frac{(N+M)^2}{10\kappa^2},\frac{(N+M)^2}{2}\right]$. That is to say, $\alpha_j=\alpha_{\min}+\frac{(j-1)(\alpha_{\max}-\alpha_{\min})}{L-1}$, for $j=1,\cdots,L$.

In the following, we present an efficient quantum algorithm to choose $\widehat{\alpha}$. Taking full advantage of quantum parallelism, our quantum algorithm can efficiently estimate $E(\alpha)$ for a given candidate $\alpha$. Since the algorithm is inspired by above $K$-fold cross-validation,  we name it \emph{quantum $K$-fold cross-validation}.

Given a certain $\alpha$,  $E_1(\alpha)=\sum_{l=1}^K \norm{\mathbf{y}_l}^2=\norm{\mathbf{y}}^2$ can be estimated easily as shown in appendix \ref{appendix:StatePre}. To estimate $E_2(\alpha)$ and $E_3(\alpha)$, $\mathbf{w}_{l}$ for $l=1,\cdots,K$ need to be revealed. Moreover, every data point in the $l$-th subset is assigned to the same $\mathbf{w}_{l}$. Therefore, we intend to generate the quantum state approximating
\begin{eqnarray}
|\psi_\mathbf{w}\rangle = \frac{\sum_{l=1}^K\left(\sum_{\tau\in S_l}|\tau\rangle\right)\otimes \mathbf{w}_l}{\sqrt{\sum_{l=1}^KN\norm{\mathbf{w}_l}^2/K}}
\label{alg2:DesState}
\end{eqnarray}
which encodes $\mathbf{w}_l$ in parallel, within error $\epsilon$.

The details of the second algorithm are described in the following steps.

(1) Prepare the initial quantum state
\begin{eqnarray}
|\psi_0\rangle=\frac{\sum_{l=1}^K \left(\sum_{\tau\in S_l}|\tau\rangle\right) \otimes \norm{\mathbf{y}_{-l}}|0,\mathbf{y}_{-l}\rangle}{\sqrt{\sum_{l=1}^KN\norm{\mathbf{y}_{-l}}^2/K}},
\label{alg2:ParallelY}
\end{eqnarray}
which can be efficiently generated in time $O(\polylog N)$ as shown in appendix \ref{appendix:StatePre}.

(2) Perform phase estimation on the above state by simulating the unitary operation
\begin{eqnarray}
\sum_{l=1}^K\left(\sum_{\tau\in S_{l}}|\tau\rangle \langle \tau|\right) \otimes e^{-\frac{i\tilde{\mathbf{X}}_{-l}t_2}{N+M}}
\label{alg2:ConSim}
\end{eqnarray}
for some evolution time $t_2$ to reveal the eigenvalues of $\frac{\tilde{\mathbf{X}}_{-l}}{N+M}$ in parallel, where
\begin{eqnarray}
\tilde{\mathbf{X}}_{-l}=\left[
\begin{matrix}
0 &\mathbf{X}_{-l} \\
\mathbf{X}_{-l}^T &0
\end{matrix}
\right]\in \mathbb{R}^{(N+M)\times(N+M)}
\end{eqnarray}
which has eigenvalues $\{\pm\lambda_{lj}\}_{j=1}^{R_l}$ and corresponding eigenvectors $\{|\mathbf{u}_{lj},\pm \mathbf{v}_{lj}\rangle\}_{j=1}^{R_l}$. Similar to the state (\ref{alg1:PE}), the resultant state becomes
\begin{eqnarray}
\frac{\sum\limits_{l=1}^K \left(\sum\limits_{\tau\in S_l}|\tau\rangle\right)\left(\sum\limits_j\norm{\mathbf{y}_{-l}}\beta_{lj}|\mathbf{u}_{lj},\pm \mathbf{v}_{lj}\rangle\ket{\frac{\pm\lambda_{lj}}{N+M}} \right) }{\sqrt{\sum_{l=1}^K 2 N\norm{\mathbf{y}_{-l}}^2/K}},
\label{alg2:PE}
\end{eqnarray}
where $\beta_{lj}:=\langle 0,\mathbf{u}_{lj}|0,\mathbf{y}_{-l}\rangle$.

(3) Similar to step (3) of algorithm 1, an auxiliary qubit is added and rotated from $|0\rangle$ to  $\sqrt{1-C_2^2h^2(\pm\lambda_{lj},\alpha)}|0\rangle + C_2h(\pm\lambda_{lj},\alpha)|1\rangle$. Here $C_2=O\left(\max_{\lambda}h(\lambda,\alpha)\right)^{-1}=O(1/\kappa')$ with $\frac{\lambda}{N+M} \in [\frac{1}{\kappa'},1]$.

(4) Undo phase estimation and measure the auxiliary qubit to see the outcome $|1\rangle$ with probability
\begin{align}
P_{\mathbf{w}}&=\frac{\sum_{l=1}^K\sum_jC_2^2\beta_{lj}^2h^2\left(\lambda_{lj},\alpha\right)\norm{\mathbf{y}_{-l}}^2}{\sum_{l=1}^K\norm{\mathbf{y}_{-l}}^2} \nonumber\\
&= \frac{\sum_{l=1}^KC_2^2(N+M)^2\norm{\mathbf{w}_l}^2}{(K-1)\norm{\mathbf{y}}^2},
\end{align}
which scales as $\Omega(1/\kappa'^2\kappa^2)$ as shown in appendix \ref{appendix:ScaPw}. To reduce the complexity, amplitude amplification is applied with $O(\kappa'\kappa)$ repetitions. Then we get the desired state $|\psi_\mathbf{w}\rangle$ of \eqref{alg2:DesState}.

(5) Append two additional registers $\ket{0\cdots 0}\ket{0}$ to the state $|\psi_\mathbf{w}\rangle$ which can be rewritten as
\begin{eqnarray}
|\psi_\mathbf{w}\rangle=\frac{\sum_{l=1}^K\left(\sum_{\tau\in S_l}|\tau\rangle\right)\otimes(\sum_{k=1}^M \mathbf{w}_{lk}|k\rangle)}{\sqrt{\sum_{l=1}^KN\norm{\mathbf{w}_l}^2/K}},
\end{eqnarray}
where $\mathbf{w}_{lk}$ is the $k$th entry of $\mathbf{w}_{l}$. Then implement the following procedures.

First, perform $O_{\mathbf{X}}$ to implement
$$\sum_{\tau\in S_l}\mathbf{w}_{lk}|\tau\rangle|k\rangle\ket{0\cdots 0}\ket{0} \mapsto \sum_{\tau\in S_l}\mathbf{w}_{lk}|\tau\rangle|k\rangle|x_{\tau k}\rangle\ket{0}.$$

Second, perform a controlled rotation denoted by controlled-$R_{\mathbf{X}}$ to generate
\begin{eqnarray*}
\sum_{\tau\in S_l}\mathbf{w}_{lk}|\tau\rangle|k\rangle|x_{\tau k}\rangle\left(\frac{x_{\tau k}}{\norm{\mathbf{X}}_{\max}}|1\rangle
+\sqrt{1-\frac{x_{\tau k}^2}{\norm{\mathbf{X}}_{\max}^2}}|0\rangle \right).
\end{eqnarray*}

Third, perform the inverse of $O_{\mathbf{X}}$ and the state becomes
\begin{eqnarray*}
 \sum_{\tau\in S_l}\mathbf{w}_{lk}|\tau\rangle|k\rangle\left(\sqrt{1-\frac{x_{\tau k}^2}{\norm{\mathbf{X}}_{\max}^2}}|0\rangle + \frac{x_{\tau k}}{\norm{\mathbf{X}}_{\max}}|1\rangle\right).
\end{eqnarray*}

Finally, perform the projective measurement on the last two registers to see if they are in the state $\left(\frac{\sum_{k=1}^M|k\rangle}{\sqrt{M}}\right)\ket{1}$, and if success we get the state (of the first register)
\begin{eqnarray}
\label{haty}
|\hat{\mathbf{y}}\rangle
&=&\frac{\sum_{l=1}^K\sum_{\tau\in S_l}(\sum_{k=1}^M\mathbf{w}_{lk}x_{\tau k})|\tau\rangle}{\sqrt{\sum_{l=1}^K\sum_{\tau\in S_l}(\sum_{k=1}^M\mathbf{w}_{lk}x_{\tau k})^2}}\nonumber\\
&=&\frac{\sum_{l=1}^K\sum_{\tau\in S_l}\mathbf{w}_l^T \mathbf{x}_{\tau}|\tau\rangle}{\sqrt{\sum_{l=1}^K\sum_{\tau\in S_l}(\mathbf{w}_l^T\mathbf{x}_{\tau})^2}}
\end{eqnarray}
encoding the prediction of $\mathbf{y}$. The success probability is
\begin{eqnarray}
P_1=\frac{\sum_{l=1}^K\sum_{\tau\in S_l}(\mathbf{w}_l^T \mathbf{x}_{\tau})^2}{M\norm{X}_{\max}^2(\sum_{l=1}^KN\norm{\mathbf{w}_l}^2/K)},
\end{eqnarray}
which, as shown in appendix \ref{appendix:ScaP1P2}, scales as $\Omega(1/\kappa'^2)$ when RR achieves good predictive performance.
This implies 
\begin{align}
\label{eq:secondterm}
E_2(\alpha)&=\sum_{l=1}^K\norm{\mathbf{X}_l\mathbf{w}_l}^2\nonumber\\
&=\sum_{l=1}^K\sum_{\tau\in S_l}(\mathbf{w}_l^T\mathbf{x}_\tau)^2 \nonumber\\
&=\frac{P_1P_\mathbf{w}NM(K-1)\norm{\mathbf{X}}_{\max}^2\norm{\mathbf{y}}^2}{C_2^2(N+M)^2K}.
\end{align}
 Noting that $N,M,K$, $\norm{\mathbf{X}}_{\max}$ and $C_2$ are known, and $\norm{\mathbf{y}}^2$ can be estimated as shown in appendix \ref{appendix:StatePre}, we can estimate $E_2(\alpha)$ by estimating $P_1$ and $P_\mathbf{w}$.

(6) Perform a swap test \cite{BCWW01,LMR13} on the states $|\mathbf{y}\rangle$ and $|\hat{\mathbf{y}}\rangle$, with the success probability of getting $\ket{0}$ being
\begin{eqnarray}
\label{eq:P2}
P_2
&=&\frac{1}{2}+\frac{1}{2}|\langle \mathbf{y}|\hat{\mathbf{y}}\rangle |^2 \nonumber\\
&=&\frac{1}{2}+\frac{1}{2}\frac{(\sum_{l=1}^K\sum_{\tau\in S_l}y_\tau\mathbf{w}_l^T \mathbf{x}_{\tau})^2}{\norm{\mathbf{y}}^2(\sum_{l=1}^K\sum_{\tau\in S_l}(\mathbf{w}_l^T\mathbf{x}_{\tau})^2)}.
\end{eqnarray}
So we have
\begin{align}
\label{eq:E3alphaEst}
&E_3(\alpha)=\sum_{l=1}^K -2\mathbf{y}_l^T\mathbf{X}_l\mathbf{w}_l
=\sum_{l=1}^K\sum_{\tau\in S_l}y_\tau\mathbf{x}_\tau^T\mathbf{w}_l \nonumber\\
&=\pm\sqrt{\frac{(2P_2-1)P_1P_{\mathbf{w}}NM(K-1)}{K}}\frac{\norm{\mathbf{y}}^2\norm{\mathbf{X}}_{\max}}{C_2(N+M)}\nonumber\\
&= \pm 2 \sqrt{(2P_2-1)E_1(\alpha)E_2(\alpha)},
\end{align}
which is ambiguous in the sign. A more deliberate method revealing the sign is to conditionally prepare these two states to make them entangled with an ancilla qubit, $\frac{|0\rangle|\mathbf{y}\rangle+|1\rangle|\hat{\mathbf{y}}\rangle}{\sqrt{2}}$, and perform the swap test on the ancilla qubit with $\frac{|0\rangle-|1\rangle}{\sqrt{2}}$ \cite{SSP16,LMR13}. The success probability is $1-\langle \mathbf{y} |\hat{\mathbf{y}}\rangle $, which reveals the exact value of $\langle \mathbf{y} |\hat{\mathbf{y}}\rangle $.
In fact, when the RR model is well constructed, the predictive outputs $\mathbf{X}_l\mathbf{w}_l$ should be close to the actual outputs $\mathbf{y}_l$, for $l=1,\cdots,K$, thus in this case the sum of their inner products, $\sum_{l=1}^K \mathbf{y}_l^T\mathbf{X}_l\mathbf{w}_l$, will be positive. Now that all three terms of Eq.~\eqref{eq:SRS} can be estimated, thus $E(\alpha)$ (eq.\eqref{alg2:Ealpha}) can be estimated as well.

(7) For every $\alpha \in \{\alpha_1, \cdots, \alpha_L\}$, execute steps (1)-(6), and then pick out the best $\alpha$ with minimum $E(\alpha)$ as the final regularization hyperparameter $\hat{\alpha}$ for RR.

\begin{figure}[hbt]
\begin{center}
\includegraphics[scale=0.36]{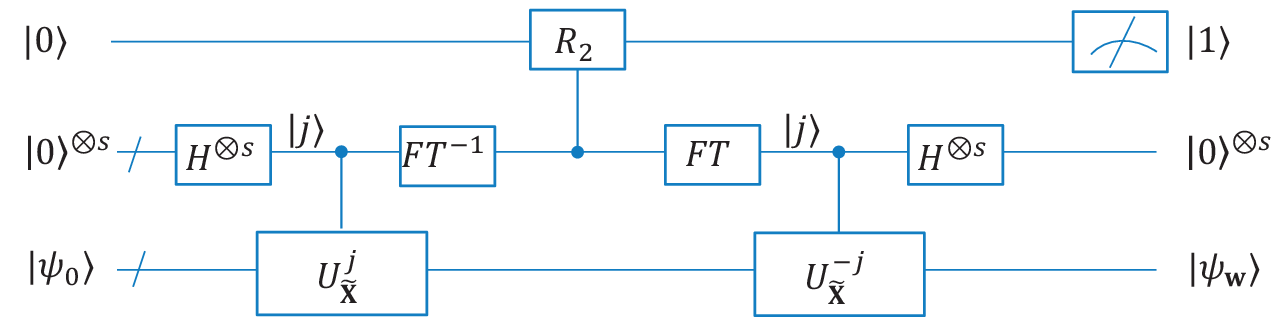}
\end{center}
\caption{\label{fig2} Quantum circuit for steps (1)-(4) of algorithm 2. Here $U_X=\sum_{l=1}^K\left(\sum_{\tau\in S_{l}}|\tau\rangle \langle \tau|\right) \otimes e^{-\frac{i\tilde{\mathbf{X}}_{-l}t_2}{(N+M)2^s}}$ .}
\end{figure}

\begin{figure}[hbt]
\begin{center}
\includegraphics[scale=0.36]{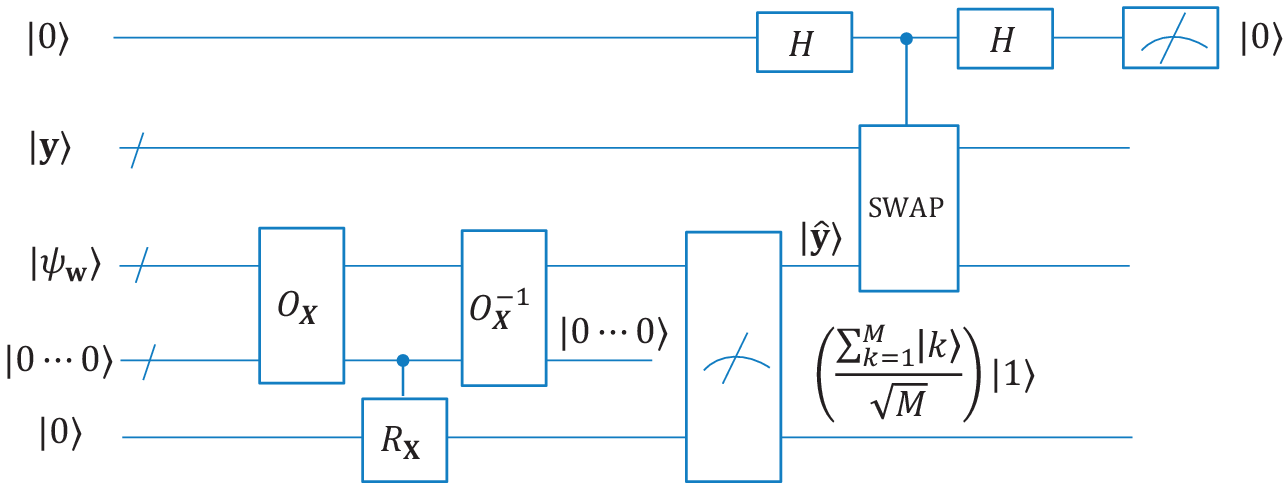}
\end{center}
\caption{\label{fig3} Quantum circuit for steps (5)-(6) of algorithm 2. Here SWAP denotes the SWAP operation.}
\end{figure}

The schematic quantum circuit of steps (1)-(4) of algorithm 2 is given in Fig. \ref{fig2} and that of steps (5)-(6) is shown in Fig. \ref{fig3}

In order to implement the unitary operation~\eqref{alg2:ConSim} for phase estimation in step (2), we propose the technique of \textit{parallel Hamiltonian simulation}, which is to simulate a chain of $N\times N$ Hermitian matrices $A_1,\cdots,A_Q$ in quantum parallel, i.e., to implement the unitary operation $\sum_{q=1}^Q \ketbra{q}{q}\otimes e^{-\frac{iA_qt}{N}}$, within some error. This technique is detailed by the following theorem whose proof is given in appendix \ref{appendix:ProofTh1}.

\begin{theorem}
\label{Thrm:ParHamSim}
(Parallel Hamiltonian simulation) Given $Q$ Hermitian $N\times N$ matrices (Hamiltonians) $\mathbf{A}_1,\cdots,\mathbf{A}_Q$ and efficient quantum oracles that can access the elements of these matrices, the unitary operation $\sum_{q=1}^Q|q\rangle\langle q|\otimes e^{-i\frac{\mathbf{A}_q}{N}t}$ can be simulated for time $t$ within spectral-norm error $\epsilon$ in time $O\left(M_\mathbf{A}^2t^2\polylog\left(N^2Q\right)/\epsilon\right)$, where the states $|q\rangle$ are the $Q$ computational basis states of a $Q$-dimensional quantum system and $M_\mathbf{A}$ is the maximum absolute value of all the elements of these matrices.
\end{theorem}

It is worth noting that the method for implementing parallel Hamiltonian simulation in Theorem \ref{Thrm:ParHamSim} is much more advantageous than the intuitive method: put $A_1,\cdots,A_Q$ into the diagonal of a larger matrix times $Q$, $A=\sum_{q=1}^Q \ketbra{q}{q}\otimes QA_q$, which is of size $NQ\times NQ$, and then simulate $A$ via the indefinite non-sparse Hamiltonian simulation \cite{RSML18}, i.e., implement the desired unitary operation $e^{-\frac{iAt}{NQ}}=\sum_{q=1}^Q\ketbra{q}{q}\otimes e^{-\frac{iA_qt}{N}}$ within error $\epsilon$. However, since $\norm{A}_{\max}=QM_\mathbf{A}$, the time complexity of this intuitive method is $O\left(Q^2M_\mathbf{A}^2t^2\polylog\left(NQ\right)/\epsilon\right)$, which is roughly $Q^2$ times more than that of the method presented in Theorem \ref{Thrm:ParHamSim}. Therefore, our method in Theorem \ref{Thrm:ParHamSim} is much more efficient than the intuitive method, especially when $Q$ is large.

According to Theorem \ref{Thrm:ParHamSim}, by setting $Q=N$, $M_\mathbf{A}=\norm{\mathbf{X}}_{\max}$ and $A_q=\tilde{\mathbf{X}}_{-l}$ for any $q\in S_l$, the unitary operation \eqref{alg2:ConSim} can be implemented within error $\epsilon$ in time $O(\norm{\mathbf{X}}_{\max}^2\polylog(N+M)t^2/\epsilon)$.

\subsection{Time complexity of algorithm 2.}

In steps (1)-(4), the time cost is mainly taken for phase estimation and amplitude amplification for generating the state $|\psi_\mathbf{w}\rangle$. Similar to algorithm 1, in step (2), $t_2=O(\kappa'/\epsilon)$ is required to make the error of $|\psi_\mathbf{w}\rangle$ be within $\epsilon$, and thus phase estimation takes time $O(\norm{\mathbf{X}}_{\max}^2\polylog(N+M)\kappa'^2/\epsilon^3)$. Plus amplitude amplification in step (4) with $O(\kappa'\kappa)$ repetitions, it takes total time $O(\norm{\mathbf{X}}_{\max}^2\polylog(N+M)\kappa'^3\kappa/\epsilon^3)$ to generate $|\psi_\mathbf{w}\rangle$.

In step (5), $E_2(\alpha)$ can be estimated by estimating $P_{\mathbf{w}}$ and $P_1$ as shown in Eq.~\eqref{eq:secondterm}. Just as estimating $\norm{\mathbf{y}}^2$ by amplitude estimation described in appendix \ref{appendix:StatePre}, $P_{\mathbf{w}}$ can be estimated within relative error $\epsilon_{\mathbf{w}}$ by amplitude estimation with
$O\left(\sqrt{\frac{1-P_{\mathbf{w}}}{P_{\mathbf{w}}}}\frac{1}{\epsilon_{\mathbf{w}}}\right)=O\left(\frac{1}{\sqrt{P_{\mathbf{w}}}\epsilon_{\mathbf{w}}}\right)$
repetitions of steps (1)-(3), resulting in the runtime
\begin{eqnarray}
O\left(\frac{\norm{\mathbf{X}}_{\max}^2\polylog(N+M)\kappa'^2/\epsilon^3}{\sqrt{P_{\mathbf{w}}}\epsilon_{\mathbf{w}}}\right).
\end{eqnarray}
Similarly, $P_1$ can be estimated within relative error $\epsilon_1$ by amplitude estimation with $O\left(\frac{1}{\sqrt{P_1}\epsilon_1}\right)$ repetitions of generating $\ket{\psi_\mathbf{w}}$ and calling $O_{\mathbf{X}}$ and $O_{\mathbf{X}}^{-1}$ as shown in step (5), which results in the runtime
\begin{eqnarray}
O\left(\frac{\norm{\mathbf{X}}_{\max}^2\polylog(N+M)\kappa'^3\kappa/\epsilon^3}{\sqrt{P_1}\epsilon_1}\right).
\end{eqnarray}
It should be noted that it is hard to estimate the scale of $P_1$ (and $P_2$) because it depends on the closeness between the prediction $\mathbf{w}_l^T \mathbf{x}_{\tau}$ and the actual output $y_\tau$ for $l=1,\cdots,K$ and $\tau\in S_l$. But when RR achieves good predictive performance with $\mathbf{w}_l^T \mathbf{x}_{\tau}\approx y_\tau$ for most $\tau$, $P_2\approx 1$, and $P_1=\Omega(1/\kappa'^2)$ as shown in appendix \ref{appendix:ScaP1P2}. In this case, $E_3(\alpha)$ would be negative and 
\begin{align}
\label{eq:ReducedSRS}
E(\alpha)&=1-\frac{2\sqrt{(2P_2-2)E_1(\alpha)E_2(\alpha)}}{E_1(\alpha)+E_2(\alpha)} \nonumber\\
&=1-\frac{2\sqrt{(2P_2-2)}\sqrt{E_2(\alpha)/E_1(\alpha)}}{1+E_2(\alpha)/E_1(\alpha)},
\end{align}
according to the eqs.\eqref{alg2:Ealpha} and \eqref{eq:E3alphaEst}. Moreover, the relative errors for estimating $P_\mathbf{w}$ and $P_1$, i.e., $\epsilon_{\mathbf{w}}$ and $\epsilon_1$, make the relative error of estimating $P_1P_\mathbf{w}$ and $E_2(\alpha)/E_1(\alpha)$ shown in Eq.~\eqref{eq:secondterm}, both be within $O(\epsilon_{\mathbf{w}} + \epsilon_1)$. Note that $E_1(\alpha)=\sum_{l=1}^K \norm{\mathbf{y}_l}^2=\norm{\mathbf{y}}^2$. Consequently, $E_2(\alpha)/E_1(\alpha)$ can be estimated within relative error $O(\epsilon_{\mathbf{w}} + \epsilon_1)$ in total time
\begin{align}
&O\left[\frac{\norm{\mathbf{X}}_{\max}^2\polylog(N+M)\kappa'^2}{\epsilon^3}
\left(\frac{1}{\sqrt{P_\mathbf{w}}\epsilon_\mathbf{w}}+\frac{\kappa'\kappa}{\sqrt{P_1}\epsilon_1}\right)\right]\nonumber\\
&=O\left[\frac{\norm{\mathbf{X}}_{\max}^2\polylog(N+M)\kappa'^3\kappa}{\epsilon^3}
\left(\frac{1}{\epsilon_\mathbf{w}}+\frac{\kappa'}{\epsilon_1}\right)\right],
\end{align}
since $P_{\mathbf{w}}=\Omega(1/\kappa'^2\kappa^2)$ and $P_1=\Omega(1/\kappa'^2)$.

In step (6), $P_2$ can be estimated within relative error $\epsilon_2$ by amplitude estimation with $O\left(\frac{1}{\sqrt{P_2}\epsilon_2}\right)$ repetitions of generating $\ket{\mathbf{y}}$ and $\ket{\hat{\mathbf{y}}}$. The state $\ket{\mathbf{y}}$ is generated in time $O(\polylog N)$ as shown in appendix \ref{appendix:StatePre}. With help of amplitude amplification, generating $\ket{\hat{\mathbf{y}}}$ in step (5) takes time $O\left(\frac{1}{\sqrt{P_1}}\frac{\norm{\mathbf{X}}_{\max}^2\polylog(N+M)\kappa'^3\kappa}{\epsilon^3}\right)$. As a result, $P_2$ can be estimated within relative error $\epsilon_2$ with runtime $O\left(\frac{\norm{\mathbf{X}}_{\max}^2\polylog(N+M)\kappa'^4\kappa}{\epsilon^3\epsilon_2}\right)$ since $P_1=\Omega(1/\kappa'^2)$ and $P_2\approx 1$ as shown in appendix \ref{appendix:ScaP1P2}. This makes the relative error for estimating $\sqrt{2P_2-1}$ be within $O(\epsilon_2)$ due to $P_2 \approx 1$. In addition, the relative error of estimating $\frac{\sqrt{E_2(\alpha)/E_1(\alpha)}}{1+E_2(\alpha)/E_1(\alpha)}$ scales as $O(\epsilon_{\mathbf{w}} + \epsilon_1)$ due to the relative error of estimating $E_2(\alpha)/E_1(\alpha)$ is also within $O(\epsilon_{\mathbf{w}} + \epsilon_1)$. According to the eq.\eqref{eq:ReducedSRS}, estimating $\frac{2\sqrt{(2P_2-2)}\sqrt{E_2(\alpha)/E_1(\alpha)}}{1+E_2(\alpha)/E_1(\alpha)}$ would incur absolute error 
\begin{align}
&O\left(\frac{2\sqrt{(2P_2-2)}\sqrt{E_2(\alpha)/E_1(\alpha)}}{1+E_2(\alpha)/E_1(\alpha)}(\epsilon_2+\epsilon_{\mathbf{w}} + \epsilon_1)\right) \nonumber\\
&=O(\epsilon_2+\epsilon_{\mathbf{w}} + \epsilon_1), \nonumber
\end{align}
as $\frac{2\sqrt{(2P_2-2)}\sqrt{E_2(\alpha)/E_1(\alpha)}}{1+E_2(\alpha)/E_1(\alpha)} \le 1$ always holds, thus $E(\alpha)$ can be estimated within absolute error $O(\epsilon_2+\epsilon_{\mathbf{w}} + \epsilon_1)$ in time $O\left(\frac{\norm{\mathbf{X}}_{\max}^2\polylog(N+M)\kappa'^4\kappa}{\epsilon^3\epsilon_2}\right)$.

Note that $E_1(\alpha)$ can be estimated within relative error $\epsilon_{\mathbf{y}}$ taking runtime $O\left(\polylog(N)/\epsilon_{\mathbf{y}}\right)$ as shown in appendix \ref{appendix:StatePre}. Letting $\epsilon_{\mathbf{y}}=\epsilon_\mathbf{w}=\epsilon_1=\epsilon_2=\epsilon$, $E(\alpha)$ can be estimated within absolute error $O(\epsilon)$, and the runtime scales as $O\left(\frac{\norm{\mathbf{X}}_{\max}^2\polylog(N+M)\kappa'^4\kappa}{\epsilon^4}\right)$. Furthermore, step (7) involves estimating $E(\alpha_1),\cdots,E(\alpha_L)$, so algorithm 2 takes runtime
\begin{align}
\label{alg2:runtime}
O\left(\frac{L\norm{\mathbf{X}}_{\max}^2\polylog(N+M)\kappa'^4\kappa}{\epsilon^4}\right)
\end{align}
in total.

The best classical counterpart of algorithm 2 consists of $L$ iterations and each one evolves a $K$-fold cross validation. In $j$th ($j=1,2,\cdots,L$) iteration, two phases are involved: (1) $K$ turns of RR with $\alpha_j$ are run and in $l$th turn $\mathbf{w}_l$ (Eq.~\eqref{alg2:DesClaL}) is output in time $O\left(\frac{(K-1)NM}{K}+\frac{(K-1)^2N^2R_l\log(R_l/\epsilon)}{K^2\epsilon^2}\right)$ \cite{CLLKZ15}; (2) $E(\alpha_j)$ is calculated according to Eq.~\eqref{eq:SRS}, which is easy to see the time complexity scales as $O(NM)$. So the total time complexity scales as $O\left(LNM+ \frac{LN^2(\sum_{l=1}^KR_l\log(R_l/\epsilon))}{\epsilon^2}\right)$ due to $K \geq 2$.

Considering the assumptions of $\norm{\mathbf{X}}_{\max}=\Theta(1)$ and $N=\Theta(M)$, and $\kappa'=O(\kappa)$ by setting $K=\Omega(\frac{NM\norm{\mathbf{X}}_{\max}^2\kappa^2}{(N+M)^2})=O(\kappa^2)$, our algorithm 2 has runtime $O(L\polylog(N)\kappa^5/\epsilon^4)$, while the best classical counterpart has runtime
$\tilde{O}\left[LN^2\left(\sum_{l=1}^K R_l\right)/\epsilon^2\right]$,
where $\tilde{O}$ is used to suppress the relatively small quantities of $\log(R_l/\epsilon)$.
When $\kappa=O(\sqrt{N})$ which is amenable to full or approximately full rank of $\mathbf{X}$ and $\mathbf{X}_{-l}$, i.e., $R,R_l=O(N)$, our algorithm 2 has runtime $O(L\polylog(N)N^{2.5}/\epsilon^4)$, whereas the classical counterpart takes time $O(L\polylog(N)N^4/\epsilon^2)$, so polynomial speedup over the classical counterpart can be achieved when $L, 1/\epsilon=O(\polylog N)$. However, when $\kappa=O(\polylog(N))$ which implies low rank of $\mathbf{X}$ as well as $\mathbf{X_{-l}}$, i.e., $R,R_l=\polylog(N)$, our algorithm 2 has runtime $O(L\polylog(N)/\epsilon^4)$, while the classical counterpart has runtime $O(L\polylog(N)N^2/\epsilon^2)$, so in this case exponential speedup can be achieved when $L, 1/\epsilon=O(\polylog N)$.

\subsection{The whole quantum algorithm for RR}

Our quantum algorithm for RR will start with algorithm 2 to find a good $\alpha$, and then plug such $\alpha$ into algorithm 1 to estimate the optimal fitting parameters in the quantum state form. The quantum state can further be applied to efficiently predict new data via swap test. It is easy to see the time complexity of the whole algorithm is dominated by algorithm 2, and thus the speedup over the classical algorithm also depends on the condition number of design matrix as discussed in algorithm 2 above.

\section{Conclusions}
In summary, we have described a quantum algorithm that can efficiently implement RR over an exponentially large data set. In particular, we propose the technique of parallel Hamiltonian simulation and use it to develop the quantum $K$-fold cross-validation that can efficiently evaluate the predictive performance of RR. The algorithm first uses quantum $K$-fold cross-validation to efficiently determine a good $\alpha$ with which RR can achieve good predictive performance, and then generates a quantum state encoding the optimal fitting parameters of RR with such $\alpha$. The state can be further used to efficiently predict new data. It is shown that our algorithm can handle data sets with non-sparse design matrices, and is able to be exponentially faster than the classical algorithm for (low-rank) design matrices with low condition numbers, but be polynomially faster than the classical algorithm for (full or approximately full) design matrices with large condition numbers.

We hope our algorithm and especially the key techniques used in our algorithm, parallel Hamiltonian simulation and quantum $K$-fold cross-validation, can inspire more efficient quantum machine learning algorithms. For example, since cross-validation is an important technique being widely used to estimate the predictive performance of various machine learning algorithms \cite{ML12,HZZL16} other than RR, it is promising that our quantum $K$-fold cross-validation can be applicable in these fields. We explore these possibilities in the future.

\ifCLASSOPTIONcompsoc
  \section*{Acknowledgments}
\else
  \section*{Acknowledgment}
\fi

We would like to thank J. B. Wang, L.-C. Wan, S.-J. Pan, H.-L. Liu, D. Li, B. J. Duan and S. Hua for helpful discussions. We also thank Sayaki Matsushita for discussions on the error analysis. This work is supported by NSFC (Grant Nos. 61572081, 61671082, and 61672110).

\ifCLASSOPTIONcaptionsoff
  \newpage
\fi

\clearpage
\setcounter{page}{1}
\appendices

\section{State preparation}
\label{appendix:StatePre}
1. Preparing the state $|\mathbf{y}\rangle$ for balanced $\mathbf{y}$.

As assumed, we are provided the quantum oracle $O_\mathbf{y}$ that can be efficiently implemented in time $O(\polylog N)$ to access the elements of $\mathbf{y}$ and acts as
\begin{eqnarray}
O_\mathbf{y}|j\rangle|0\rangle=|j\rangle|y_j\rangle.
\label{eq:Oy}
\end{eqnarray}
We start with performing the oracle on the state $\frac{\sum_{j=1}^N|j\rangle|0\rangle}{\sqrt{N}}$ to have $\sum_{j=1}^N\frac{|j\rangle|y_j\rangle}{\sqrt{N}}$. Then we append a qubit and perform controlled rotation to generate the state
\begin{eqnarray*}
\sum_{j=1}^N\frac{|j\rangle|y_j\rangle}{\sqrt{N}}
\left(\sqrt{1-\left(\frac{y_j}{\norm{\mathbf{y}}_{\max}}\right)^2}|0\rangle+\frac{y_j}{\norm{\mathbf{y}}_{\max}}|1\rangle\right).
\end{eqnarray*}
Finally, uncompute the oracle and measure the last qubit to see $|1\rangle$ with probability $P_\mathbf{y}=\frac{\sum_{j=1}^Ny_j^2}{N\norm{\mathbf{y}}^2_{\max}}$. The final state of the first register would be $|\mathbf{y}\rangle$ as desired. Since $\mathbf{y}$ is balanced, $P_\mathbf{y}=\Omega(1)$. This means we need $O(1)$ measurements (as well as oracles $O_\mathbf{y}$) to obtain $|\mathbf{y}\rangle$ with a large probability, and thus the total time for generating $|\mathbf{y}\rangle$ is $O(\polylog N)$.

In addition, $P_\mathbf{y}$ can be estimated within error $\hat{\epsilon}_{\mathbf{y}}$ by amplitude estimation \cite{AA02} using $O\left(\sqrt{P_\mathbf{y}(1-P_\mathbf{y})}/\hat{\epsilon}_{\mathbf{y}}\right)$ repetitions of $O_\mathbf{y}$ and its inverse (as required above), and thus $P_\mathbf{y}$ can be estimated within relative error $\epsilon_{\mathbf{y}}=\hat{\epsilon}_{\mathbf{y}}/P_\mathbf{y}$ in runtime
\begin{eqnarray*}
&&O\left(\sqrt{(1-P_\mathbf{y})/P_\mathbf{y}}/\epsilon_{\mathbf{y}}\times\polylog(N)\right)\nonumber\\
&=&O\left(\polylog(N)/\epsilon_{\mathbf{y}}\right)
\end{eqnarray*}
since $P_\mathbf{y}=\Omega(1)$. Moreover, since $\norm{\mathbf{y}}^2=\sum_{j=1}^N y_j^2=N P_\mathbf{y} \norm{\mathbf{y}}_{\max}$, $\norm{\mathbf{y}}^2$ can be estimated by estimating $P_{\mathbf{y}}$ within relative error $\epsilon_{\mathbf{y}}$ in runtime $O\left(\polylog(N)/\epsilon_{\mathbf{y}}\right)$.

2. Preparing $|\psi_0\rangle$ (initial state of algorithm 2).

To generate the state
\begin{eqnarray}
\frac{\sum_{l=1}^K \left(\sum_{\tau\in S_l}|\tau\rangle\right) \otimes \norm{\mathbf{y}_{-l}}|0,\mathbf{y}_{-l}\rangle}{\sqrt{\sum_{l=1}^KN\norm{\mathbf{y}_{-l}}^2/K}}\nonumber\\
=\frac{\sum_{l=1}^K \left(\sum_{\tau\in S_l}|\tau\rangle\right) \otimes \left(\sum_{j=1,j\notin S_l}^N \mathbf{y}_j|0,j\rangle\right)}{\sqrt{\sum_{l=1}^KN\norm{\mathbf{y}_{-l}}^2/K}},
\label{IniStateOfalg2}
\end{eqnarray}
we first prepare
\begin{eqnarray}
&&  \left(\frac{\sum_{i=1}^N|i\rangle}{\sqrt{N}}\right)|0,\mathbf{y}\rangle \nonumber\\
&=& \left(\frac{\sum_{l=1}^K\sum_{\tau \in S_l}|\tau\rangle}{\sqrt{N}}\right)
\left(\frac{\sum_{j=1}^N \mathbf{y}_j|0,j\rangle}{\norm{\mathbf{y}}}\right),
\label{eq:yl}
\end{eqnarray}
where $|0,\mathbf{y}\rangle$ is the $(N+M)$-dimensional state vector by adding $M$ zero entries to the state vector $|\mathbf{y}\rangle$ and $|0,j\rangle$ are the computational basis states of a $(M+N)$-dimensional quantum system. Since $|\mathbf{y}\rangle$ can be efficiently generated in time $O(\polylog N)$ as shown above, $|0,\mathbf{y}\rangle$ can be efficiently prepared as well. By comparing the states (\ref{IniStateOfalg2}) and (\ref{eq:yl}), we can find the state (\ref{IniStateOfalg2}) is the normalized vector of (\ref{eq:yl}) after kicking out the terms $\sum_{l=1}^K\left(\frac{\sum_{\tau \in S_l}|\tau\rangle}{\sqrt{N}}\right)
\left(\frac{\sum_{j\in S_l} \mathbf{y}_j|0,j\rangle}{\norm{\mathbf{y}}}\right)$. The squared amplitudes sum of the remaining terms is $\frac{K-1}{K}\geq \frac{1}{2}$ due to $K\geq 2$. This implies we can easily obtain the state (\ref{IniStateOfalg2}) from the state (\ref{eq:yl}) by adding an auxiliary qubit to state (\ref{eq:yl}) to mark the terms of state (\ref{IniStateOfalg2}) in state (\ref{eq:yl}) and measuring this qubit with probability $\frac{K-1}{K}$. Therefore, as the state (\ref{eq:yl}), the state (\ref{IniStateOfalg2}) can be efficiently generated in time $O(\polylog N)$.

\section{The maximum value and the maximum relative error of $h\left(\lambda_j,\alpha\right)$ depending on $\alpha$}
\label{appendix:MaxOfh}

The maximum value and the maximum relative error of $h\left(\lambda_j,\alpha\right)=\frac{(N+M)\lambda_j}{\lambda_j^2+\alpha}$ ($\alpha > 0 ,j=1,2,\cdots,R$) with $\lambda_j\in [\frac{N+M}{k},N+M]$ respectively determines $C_1$  and the error of final desired state of algorithm 1, and their scales depend on the actual choice of $\alpha$.  The results can also be applied to algorithm 2.

1. \emph{The maximum value of $h\left(\lambda_j,\alpha\right)$ depending on $\alpha$}.

Let us first define $h\left(\lambda,\alpha\right)=\frac{(N+M)\lambda}{\lambda^2+\alpha}$ with $\lambda \in [\frac{N+M}{k},N+M]$ and $\alpha > 0$. Its derivative on $\lambda$
\begin{eqnarray}
h'\left(\lambda,\alpha\right) = \frac{(N+M)(\alpha-\lambda^2)}{(\lambda^2+\alpha)^2}
\label{eq:DerivativeOfh}
\end{eqnarray}
implies that
\begin{eqnarray}
&&\max_{\lambda}h\left(\lambda,\alpha\right)=\nonumber\\
&&\begin{cases}
\frac{(N+M)^2\kappa}{(N+M)^2+\kappa^2\alpha} & \mathrm{when}\ \alpha \leq \frac{(N+M)^2}{\kappa^2} \\
\frac{(N+M)}{2\sqrt{\alpha}} &                 \mathrm{when}\ \frac{(N+M)^2}{\kappa^2} < \alpha \leq (N+M)^2 \\
\frac{(N+M)^2}{(N+M)^2+\alpha} &               \mathrm{when}\ (N+M)^2 < \alpha.
\end{cases}\nonumber
\label{eq:Maxh}
\end{eqnarray}
These equations for different cases of $\alpha$ give tighter and more practical upper bounds for the maximum value of $h\left(\lambda_j,\alpha\right)$, as well as the choice of $C_1$ in step (3) of algorithm 1. In addition, the facts that
\begin{eqnarray}
\frac{\max\limits_{\lambda}h\left(\lambda,\alpha\right)}{\min\limits_{\lambda}h\left(\lambda,\alpha\right)}
&=& \max\limits_{\lambda_1,\lambda_2} \frac{h\left(\lambda_1,\alpha\right)}{h\left(\lambda_2,\alpha\right)}
= \frac{\lambda_1(\lambda_2^2+\alpha)}{\lambda_2(\lambda_1^2+\alpha)} \nonumber\\
&\le& \frac{\lambda_1}{\lambda_2}\left(\frac{\lambda_2^2}{\lambda_1^2}+1\right)
= \frac{\lambda_2}{\lambda_1} +\frac{\lambda_1}{\lambda_2}
\end{eqnarray}
reaches its maximum $\le \kappa+\frac{1}{\kappa}=O(\kappa)$ when $\{\lambda_1,\lambda_2\}=\{N+M,\frac{N+M}{\kappa}\}$ (for $\kappa>1$), and that $C_1=O\left(\mathop{\max}_{\lambda_j}h(\lambda_j,\alpha)\right)^{-1}$ makes $C_1h(\lambda_j,\alpha)=\Omega(1/\kappa)$.

2. \emph{The maximum relative error of $h\left(\lambda_j,\alpha\right)$ depending on $\alpha$}.

In fact, the maximum relative error of $h\left(\lambda_j,\alpha\right)$ scales as $O\left(|g(\lambda)|\epsilon_\lambda\right)$, where
\begin{eqnarray}
g(\lambda) = \frac{h'\left(\lambda,\alpha\right)}{h\left(\lambda,\alpha\right)}
           = \frac{\alpha-\lambda^2}{\lambda(\lambda^2+\alpha)}
\label{eq:funcg}
\end{eqnarray}
and $\epsilon_\lambda=O\left(\frac{N+M}{t}\right)$ is the estimate error for estimating $\lambda$ ($\lambda_j$) by phase estimation (step (2) of algorithm 1). Since
\begin{eqnarray}
g(\lambda)^2-\frac{1}{\lambda^2} = \frac{-4\alpha}{(\lambda^2+\alpha)^2} <0,
\end{eqnarray}
thus $|g(\lambda)|<\frac{1}{\lambda}$ and the relative error of $h\left(\lambda_j,\alpha\right)$ roughly scales as $O\left(\kappa/t\right)$ regardless of $\alpha$. To obtain the more precise and practical relative error, we take the derivative of $g^2(\lambda)$ on $\lambda$,
\begin{align}
&(g^2(\lambda))'=\nonumber\\
&\frac{2(\alpha-\lambda^2)\left(\lambda^2-(2+\sqrt{5})\alpha\right)\left(\lambda^2-(2-\sqrt{5})\alpha\right)}{\lambda^3(\lambda^2+\alpha)^3},
\label{eq:Derivativeg2}
\end{align}
which implies $\max_{\lambda}\abs{g(\lambda)}=$

\begin{enumerate}[(1)]
  \item $\frac{(N+M)^2\kappa-\kappa^3\alpha}{(N+M)\left((N+M)^2+\kappa^2\alpha\right)}$, $\alpha \in [0,\frac{(N+M)^2}{(2+\sqrt{5})\kappa^2}]$;
  \item $\frac{1+\sqrt{5}}{\sqrt{2+\sqrt{5}}(3+\sqrt{5})\sqrt{\alpha}} \approx \frac{0.3}{\sqrt{\alpha}}$,
  $\alpha \in [\frac{(N+M)^2}{(2+\sqrt{5})\kappa^2}, \frac{(N+M)^2}{\kappa^2}]$;
  \item $\max\left\{\frac{\kappa^3\alpha-(N+M)^2\kappa}{(N+M)\left((N+M)^2+\kappa^2\alpha\right)},\frac{1+\sqrt{5}}{\sqrt{2+\sqrt{5}}(3+\sqrt{5})\sqrt{\alpha}}\right\}$, $\alpha \in [\frac{(N+M)^2}{\kappa^2}, \frac{(N+M)^2}{2+\sqrt{5}}]$;
  \item $\max\left\{\frac{\kappa^3\alpha-(N+M)^2\kappa}{(N+M)\left((N+M)^2+\kappa^2\alpha\right)},\frac{(N+M)^2-\alpha}{(N+M)((N+M)^2+\alpha)}\right\}$, $\alpha \in [\frac{(N+M)^2}{2+\sqrt{5}}, (N+M)^2]$;
  \item $\frac{\kappa^3\alpha-(N+M)^2\kappa}{(N+M)\left((N+M)^2+\kappa^2\alpha\right)}$, $\alpha \in [(N+M)^2,+\infty]$.
\end{enumerate}

These results for different choices of $\alpha$ give tighter upper bounds for the maximum relative error of $h\left(\lambda_j,\alpha\right)$ as well as tighter error estimate for the final state of algorithm 1.

\section{Scale of singular values of $\mathbf{X}_{-l}$}
\label{appendix:ScaSingXl}

According to the definitions in the main paper, $\mathbf{X}=(\mathbf{x}_1,\cdots,\mathbf{x}_N)$ and $\mathbf{X}_{-l}$ is the matrix constructed by replacing the rows $S_l$ of $\mathbf{X}$ with zeros, thus we have
\begin{eqnarray}
\mathbf{X}^T\mathbf{X}=\mathbf{X}_{-l}^T\mathbf{X}_{-l}+\sum_{j\in S_l}\mathbf{x}_j^T\mathbf{x}_j.
\label{eq:SSlDifference}
\end{eqnarray}
Noting that the rank of $\mathbf{X}_{-l}^T\mathbf{X}_{-l}$ is evidently equal to or less than that of $\mathbf{X}^T\mathbf{X}$, the eigenvalues of $\mathbf{X}^T\mathbf{X}$ and $\mathbf{X}_{-l}^T\mathbf{X}_{-l}$ are $0<\lambda_R^2 \leq \cdots \leq \lambda_1^2$ and $0\leq\lambda_{lR}^2 \leq \cdots \leq \lambda_{l1}^2$ respectively. According to Weyl's inequality \cite{Weyl93},
\begin{eqnarray}
\lambda_j^2-\norm{\sum_{j\in S_l}\mathbf{x}_j^T\mathbf{x}_j}\leq \lambda_{lj}^2 \leq \lambda_j^2,
\label{eq:Weyl}
\end{eqnarray}
which implies $\lambda_{l1} \leq \lambda_1 \leq (N+M)^2$ and $\lambda_{lR} \geq \frac{(N+M)^2}{\kappa^2}-\frac{NM\norm{\mathbf{X}}_{\max}^2}{K}$. Therefore, $\lambda_{lj}\in \left[\frac{N+M}{\kappa'}, N+M\right]$ for some $\kappa'$, and we take $\kappa' = O(\kappa)$ by setting $K=\Omega\left(\frac{NM\norm{\mathbf{X}}_{\max}^2\kappa^2}{(N+M)^2}\right)=\Omega\left(\kappa^2\right)$ for $\norm{\mathbf{X}}_{\max}=\Theta(1)$ and $N=\Theta(M)$.

\section{Scale of $P_{\mathbf{w}}$}
\label{appendix:ScaPw}

In step (4) of algorithm 2, the measurement probability is
\begin{eqnarray}
P_{\mathbf{w}}=\frac{\sum_{l=1}^K\sum_jC_2^2\beta_{lj}^2h^2\left(\lambda_{lj},\alpha\right)\norm{\mathbf{y}_{-l}}^2}{\sum_{l=1}^K\norm{\mathbf{y}_{-l}}^2}.
\label{eq:Pw}
\end{eqnarray}
First, as proving $C_1h(\lambda_j,\alpha)=\Omega(1/\kappa)$ in appendix \ref{appendix:MaxOfh}, it is easy to prove $C_2h\left(\lambda_{lj},\alpha\right)=\Omega(1/\kappa')$. Moreover, since
\begin{eqnarray}
&&\sum_{l=1}^K\sum_j\beta_{lj}^2\lambda_{lj}^2\norm{\mathbf{y}_{-l}}^2 \nonumber\\
&& = \sum_{l=1}^K\norm{\mathbf{X}_{-l}^T\mathbf{y}_{-l}}^2  \geq  \frac{\left(\sum_{l=1}^K\norm{\mathbf{X}_{-l}^T\mathbf{y}_{-l}}\right)^2}{K} \nonumber\\
&& \geq  \frac{\norm{\sum_{l=1}^K\mathbf{X}_{-l}^T\mathbf{y}_{-l}}^2}{K}  =  \frac{(K-1)^2\norm{\mathbf{X}^T\mathbf{y}}^2}{K}\nonumber\\
&& =  \frac{(K-1)^2\sum_{j=1}^R\lambda_j^2\beta_j^2\norm{\mathbf{y}}^2}{K}\nonumber\\
&& =  \Omega\left(\frac{(N+M)^2(K-1)^2\norm{\mathbf{y}}^2}{K\kappa^2}\right)
\label{eq:Inequality}
\end{eqnarray}
and $\lambda_{lj}\leq (N+M)^2$, we can obtain
\begin{eqnarray}
\frac{\sum_{l=1}^K\sum_j\beta_{lj}^2\norm{\mathbf{y}_{-l}}^2}{\sum_{l=1}^K\norm{\mathbf{y}_{-l}}^2}&=& \frac{\sum_{l=1}^K\sum_j\beta_{lj}^2\norm{\mathbf{y}_{-l}}^2}{(K-1)\norm{\mathbf{y}}^2}\nonumber\\
&=& \Omega\left(\frac{K-1}{K\kappa^2}\right)\nonumber \\
&=& \Omega\left(\frac{1}{\kappa^2}\right)
\label{eq:ScaleOf2}
\end{eqnarray}
($K\geq 2$). Combining these two results, the scale of $P_{\mathbf{w}}$ can be derived as
\begin{eqnarray}
P_{\mathbf{w}}=\Omega\left(\frac{1}{\kappa'^2\kappa^2}\right).
\label{eq:ScaleOfPw}
\end{eqnarray}

\section{Proof of theorem 1}
\label{appendix:ProofTh1}

\begin{proof}
Our method implements simulating $\sum_{q=1}^Q|q\rangle\langle q|\otimes e^{-i\frac{\mathbf{A}_l}{N}t}$ on two quantum states $\sigma_C \otimes \sigma$ where $\sigma_C\in \mathbb{C} ^{Q\times Q}$ and $\sigma \in \mathbb{C}^{N\times N}$, assisted by multiple copies of $\rho=|\vec{1}\rangle\langle \vec{1}| \in \mathbb{C}^{N \times N}$ ($|\vec{1}\rangle=\frac{\sum_{j=1}^N|j\rangle}{\sqrt{N}}$). For simplicity but without loss of generality, we consider $\sigma_C=|q\rangle\langle q|$ for any $q\in\{1,2,\cdots,Q\}$. Similar to the indefinite density Hermitian matrix simulation \cite{RSML18}, we first imbed each Hermitian matrix $\mathbf{A}_q$ to a larger one-sparse $N^2\times N^2$ Hermitian matrix
\begin{eqnarray}
S_{\mathbf{A}_q}= \sum_{j,k=1}^N\mathbf{A}_{q,jk}|k\rangle\langle j|\otimes |j\rangle \langle k| \in \mathbb{C}^{N^2\times N^2},
\label{eq:SAl}
\end{eqnarray}
where $\mathbf{A}_{q,jk}$ are the elements of $\mathbf{A}_q$. Then the sparse matrices are imbedded to an one-sparse Hermitian matrix
\begin{eqnarray}
S_{\mathbf{A}}= \sum_{q=1}^Q|q\rangle\langle q|\otimes S_{\mathbf{A}_{q}}\in \mathbb{C}^{QN^2 \times QN^2}.
\label{eq:SA}
\end{eqnarray}
Since it is one-sparse, given the efficient quantum oracles accessing the elements of $S_{\mathbf{A}}$ that can run in time $O\left(\polylog (N^2Q)\right)$ via, for example, quantum random access memory \cite{QRAM08}, the unitary operation
\begin{eqnarray}
e^{-iS_\mathbf{A}t}= \sum_{q=1}^Q|q\rangle\langle q|\otimes e^{-iS_{\mathbf{A}_q}t}
\end{eqnarray}
for any time $t$ can be efficiently simulated with constant number of oracle calls \cite{BACS07}. Then after preparing $n=\frac{t}{\Delta t}$ copies of $\rho$, we perform $e^{-iS_\mathbf{A}t}$ on $|q\rangle \langle q|\otimes \rho\otimes \sigma$ for each copy and the resultant state of the first and third systems will become
\begin{eqnarray}
&&\Tr_2(e^{-iS_\mathbf{A}\Delta t}|q\rangle \langle q|\otimes \rho\otimes \sigma e^{iS_\mathbf{A}\Delta t})\nonumber\\
&=& |q\rangle \langle q| \otimes \left(\sigma-i\frac{\Delta t}{N}[\mathbf{A}_q,\sigma]+O(M_{\mathbf{A}_q}^2\Delta t^2)\right)\nonumber\\
&\approx & |q\rangle\langle q|\otimes e^{-i\frac{\mathbf{A}_{q}\Delta t}{N}} \sigma e^{i\frac{\mathbf{A}_{q}\Delta t}{N}}.
\end{eqnarray}
The spectral-norm error (directly implied by trace norm error in \cite{RSML18}) scales as $O\left(M_{\mathbf{A}_q}^2\Delta t^2\right)$ \cite{RSML18}, where $M_{\mathbf{A}_q}=\norm{\mathbf{A}_q}_{\max}$. Since $q\in \{1,2,\cdots,Q\}$ is arbitrary, the error should scale as $O\left(M_{\mathbf{A}}^2\Delta t^2\right)$,  where $M_\mathbf{A}$ is the maximum absolute value of the elements of all the matrices $\mathbf{A}_1,\mathbf{A}_2,\cdots,\mathbf{A}_Q$, i.e., $M_{\mathbf{A}}=\max_{q=1,\cdots,Q}M_{\mathbf{A}_q}$. Therefore, running this procedure for $n$ times allows simulating the unitary operation $\sum_{q=1}^Q|q\rangle\langle q|\otimes e^{-i\frac{\mathbf{A}_q}{N}t}$ with spectral-norm error $O\left(nM_{\mathbf{A}}^2\Delta t^2\right)$. To make the error be within $\epsilon$, $n$ should be chosen as
\begin{eqnarray}
n=O\left(\frac{M_\mathbf{A}^2t^2}{\epsilon}\right).
\label{eq:n}
\end{eqnarray}
Therefore, the total time complexity is
$$O\left(n\log(N^2Q)\right)=O\left(M_\mathbf{A}^2t^2\polylog(N^2Q)/\epsilon\right).$$
\end{proof}

\section{Scale of $P_1$ and $P_2$ when RR achieves good predictive performance}
\label{appendix:ScaP1P2}

When RR achieves good predictive performance with $\mathbf{w}_l^T \mathbf{x}_{\tau}\approx y_\tau$ for most $\tau\in S_l$ and $l=1,\cdots,K$,
\begin{eqnarray}
\label{Appeq:P1}
P_1&=&\frac{\sum_{l=1}^K\sum_{\tau\in S_l}(\mathbf{w}_l^T \mathbf{x}_{\tau})^2}{M\norm{X}_{\max}^2(\sum_{l=1}^KN\norm{\mathbf{w}_l}^2/K)}\nonumber\\
&\approx& \frac{\norm{\mathbf{y}}^2}{M\norm{X}_{\max}^2(\sum_{l=1}^KN\norm{\mathbf{w}_l}^2/K)}.
\end{eqnarray}
Moreover, since
\begin{eqnarray}
\mathbf{w}_l&=&(\mathbf{X}_{-l}^T\mathbf{X}_{-l}+\alpha \mathbf{I})^{-1}\mathbf{X}_{-l}^T\mathbf{y}_{-l} \\
&=& \sum_{j} \frac{\lambda_{lj}}{\lambda_{lj}^2+\alpha}\beta_{lj}\norm{\mathbf{y}}|\mathbf{v}_{lj}\rangle
\end{eqnarray}
and $\lambda_{lj}\in [\frac{N+M}{\kappa'},N+M]$, we have
\begin{eqnarray}
\norm{\mathbf{w}_l}^2 &=& \sum_{j} \frac{\lambda_{lj}^2}{(\lambda_{lj}^2+\alpha)^2}\beta_{lj}^2\norm{\mathbf{y}}^2 \\
&\leq& \sum_{j} \frac{\beta_{lj}^2}{\lambda_{lj}^2}\norm{\mathbf{y}}^2\\
&\leq& \sum_{j} \frac{\kappa'^2\beta_{lj}^2}{(N+M)^2}\norm{\mathbf{y}}^2\\
&=& \frac{\kappa'^2}{(N+M)^2}\norm{\mathbf{y}}^2.
\end{eqnarray}
Plugging the result to Eq.~\eqref{Appeq:P1}, we have
\begin{eqnarray}
P_1 &\ge& \frac{(N+M)^2}{MN\kappa'^2\norm{\mathbf{X}}_{\max}^2}\\
&=&\Omega(1/\kappa'^2)
\end{eqnarray}
for $M=\Theta(N)$ and $\norm{\mathbf{X}}_{\max}=\Theta(1)$.

Moreover, putting $\mathbf{w}_l^T \mathbf{x}_{\tau}\approx y_\tau$ (for every $l=1,\cdots,K$ and every $\tau\in S_l$) into Eq.~\eqref{eq:P2}, it is easy to see $P_2\approx 1$.
\end{document}